\documentclass[conference,a4paper]{IEEEtran}

\def\arxiv{1}
\usepackage[T1]{fontenc}
\usepackage[cmex10]{amsmath} % Use the [cmex10] option to ensure complicance
\usepackage[cmintegrals]{newtxmath}
\usepackage{bm} %need to manually load newtxmath before bm

\usepackage{mathtools} % for underbracket
\usepackage[shortlabels]{enumitem} %customised enumerate
\usepackage{graphicx}
\usepackage{microtype}
\usepackage[space,noadjust]{cite}
\usepackage{subcaption}
\usepackage{xcolor}
\usepackage[ruled,vlined,linesnumbered]{algorithm2e}
\SetAlFnt{\small}
\SetAlCapFnt{\small}
\SetAlCapNameFnt{\small}

\usepackage{etoolbox}
\makeatletter
\patchcmd{\@begintheorem}{\textit}{\textbf}{}{}
\patchcmd{\@begintheorem}{\itshape}{\bfseries}{}{}
\makeatother

\newcommand{\absent}{\mathbb{U}^\text{abs}}
\newtheorem{theorem}{Theorem}
\newtheorem{corollary}{Corollary}
\newtheorem{lemma}{Lemma}
\newtheorem{proposition}{Proposition}
\newtheorem{definition}{Definition}

\newtheorem{remark}{Remark}
\newtheorem{example}{Example}
\graphicspath{{.}{../pictures/}{pictures/}{../../pictures/}}

\def\gap{1.21ex}
\abovedisplayskip\gap
\belowdisplayskip\gap
\abovedisplayshortskip\gap
\belowdisplayshortskip\gap

\begin{document}

%---------- Title ----------
\title{Improved Lower Bounds for Pliable Index Coding using Absent Receivers}

\author{\IEEEauthorblockN{Lawrence Ong}
\IEEEauthorblockA{\small University of Newcastle\\
%Callaghan, NSW 2308, Australia\\
  lawrence.ong@newcastle.edu.au}
\and

\IEEEauthorblockN{Badri N.\ Vellambi}
\IEEEauthorblockA{\small University of Cincinnati\\
badri.vellambi@uc.edu}

\and

\IEEEauthorblockN{J\"{o}rg Kliewer} 
\IEEEauthorblockA{\small New Jersey Institute of Technology\\
  jkliewer@njit.edu}

\and

\IEEEauthorblockN{Parastoo Sadeghi} 
\IEEEauthorblockA{\small Australian National University\\
parastoo.sadeghi@anu.edu.au}

}

\IEEEoverridecommandlockouts
%\vspace*{-3ex}
\maketitle
%\vspace*{-4ex}

\begin{abstract}
This paper studies pliable index coding, in which a sender broadcasts information to multiple receivers through a shared broadcast medium, and the receivers each have some message a priori and want any message they do not have. An approach, based on receivers that are absent from the problem, was previously proposed to find lower bounds on the optimal broadcast rate. In this paper, we introduce new techniques to obtained better lower bounds, and derive the optimal broadcast rates for new classes of the problems, including all problems with up to four absent receivers.
\end{abstract}

%%%%
%%%%
%%%%
%%%%
% \vspace{-3ex}
\section{Introduction}

%Pliable index coding

This papers studies pliable index coding, where one transmitter sends information to multiple receivers in a noiseless broadcast setting. In the original index-coding setup~\cite{baryossefbirk11,ong2017}, each receiver is described by the set of messages that it has, referred to as side information, and the message that it wants from the transmitter. In the \textit{pliable} variant of the problem~\cite{brahmafragouli15}, each receiver is described by only its side information, and its decoding requirement is relaxed to any message not in the side-information set.

% Existing bounds
The aim for both the original and the pliable problems is to determine the minimum  codelength normalised to the message length, referred to as the optimal broadcast rate, that the transmitter must broadcast to satisfy all receivers. 
As with original index-coding problems, the optimal broadcast rate is not known for pliable-index-coding problems in general.

%\subsection{Existing results}
Even though the two index-coding versions share many similarities, their decoding requirements set them apart in non-trivial ways. As a result, different techniques have been attempted to solve each of them. To date, only a small number of classes of pliable-index-coding problems have been solved. In particular, two classes of \textit{symmetric} problems have been solved~\cite{liutuninetti17, liutuninetti18}.
%Denote the receiver that has side-information set $H$ by receiver~$H$, these problems are symmetric in the sense that all receivers with certain $|H|$'s are present, and the rest absent.
 These problems are symmetric in the sense that if a receiver is present in the problem, every receiver with the same cardinality of messages as side information as that of the present receiver is also present.
 For asymmetric problems, we derived the optimal broadcast rate for some classes of problems based on the absent receivers~\cite{ongvellambikliewer2019}.
 % Denote the receiver that has side-information set~$H$ by receiver~$H$.
 We label a receiver by its side-information set, for instance, receiver~$H$ has side-information set~$H$. With this notation, we lower bounded the optimal broadcast rate by the longest chain of nested absent receivers, that is, there exist absent receivers $H_1, H_2, \dotsc , H_{L_{\max}}$ such that $H_1\subsetneq H_2 \subsetneq \cdots \subsetneq H_{L_{\max}}$. We characterised the optimal broadcast rate when \textsf{(i)}~there exists a message not in the side-information set of any absent receiver, \textsf{(ii)}~there is no nested absent receiver pair, \textsf{(iii)}~there is only one nested absent receiver pair, and \textsf{(iv)}~the absent receivers are formed by taking unions of some message partitions.

%\subsection{Contributions}
However, with the existing results, even a simple problem with three absent receivers remained unsolved (see problem $\mathcal{P}_1$ in Section~\ref{sec:motivating-example}). In this paper, we strengthen our previous results to obtain new lower bounds. As a result of the improved lower bounds, we can solve all pliable-index-coding problems with four or fewer absent receivers (which includes $\mathcal{P}_1$).

Our previous results~\cite{ongvellambikliewer2019} were derived based on our proposed algorithm to construct a decoding chain. The algorithm iteratively adds messages to the chain.
%When the chain comprises a message set, say $H$, we add the message that receiver~$H$ decodes to the chain. If receiver~$H$ is absent, we arbitrarily ``skip'' a message by adding it to the chain.
When the current decoding chain corresponds to a present receiver~$H$, the message that receiver~$H$ wants to decode is added to the chain. If the current chain does not correspond to any present receiver, we will arbitrarily ``skip'' a message not in the chain and also add the same message to the decoding chain.
This continues till the chain equals to the whole message set. The fewer the skipped messages, the tighter the lower bound.  In this paper, we propose two improvements. First, we modify the algorithm such that even if receiver~$H$ is absent, we may not need to skip a message, by looking at receivers $H^- \subsetneq H$, and the messages to be decoded by them. Second, instead of arbitrarily skipping a message, we consider the next absent receiver $H'$ that the algorithm will encounter, and skip a message in such a way that we will be able to avoid skipping a message when the algorithm reaches $H'$.

We will formally define pliable-index-coding problems in Section~\ref{sec:formulation}, after which we will use an example to illustrate the above-mentioned two new ideas in Section~\ref{sec:motivating-example}. These two ideas will be formally presented in Sections~\ref{sec:new-algo} and \ref{sec:look-ahead}. In Section~\ref{sec:new-nested-chain}, we will also present a simpler lower bound. The results will be combined to characterise the optimal broadcast rate for new classes of pliable-index-coding problems in Section~\ref{sec:application}.

%Issues

%Solutions

\section{Problem Formulation}\label{sec:formulation}

We use the following notation: $\mathbb{Z}^+$ denotes the set of natural numbers, $[a:b] := \{a, a+1, \dotsc, b\}$ for $a,b\in\mathbb{Z}^+$ such that $a < b$, and $X_S = (X_i: i \in S)$ for some ordered set $S$.

Consider a sender having $m \in \mathbb{Z}^+$ messages, denoted by $X_{[1 : m]} = (X_1, \dots, X_m)$. Each message $X_i \in \mathbb{F}_q$  is independently and uniformly distributed over a finite field of size~$q$. There are $n$ receivers having distinct subsets of messages, which we refer to as side information. Each receiver is labelled by its side information, i.e., the  receiver that has messages $X_{H}$, for some $H \subsetneq [1 : m]$, will be referred to as receiver $H$. The aim of the pliable-index-coding problem is to devise an encoding scheme for the sender and a decoding scheme for each receiver satisfying pliable recovery of a message at each receiver. 

Without loss of generality, the side-information sets of the receivers are distinct; all receivers having the same side information can be satisfied if and only if (iff) any one of them can be satisfied. Also, no receiver has side information~$H = [1:m]$ because this receiver cannot be satisfied. So, there can be at most $2^m-1$ receivers present in the problem. A pliable index coding problem is thus defined uniquely by $m$ and the set $\mathbb{U} \subseteq 2^{[1:m]} \setminus \{[1:m]\}$ of all present receivers. %Lastly, %a receiver is said to be \textit{absent} if it is not present in the problem.
Any receiver that is not present, i.e., receiver~$H \in 2^{[1:m]} \setminus (\{[1:m]\} \cup \mathbb{U}) := \absent$, is said to be \textit{absent}.

% \begin{example}
% Let $m= 3$, and  $\mathbb U=\{\emptyset, \{1\}, \{2\}, \{1,2\}, \{2,3\}\}$. Then, the receivers $\{3\}$ and $\{1,3\}$ are absent.
% \end{example}

Given a pliable-index-coding problem with $m$ messages and present receivers $\mathbb U$, a pliable index code of length $\ell \in \mathbb{Z}^+$ consists of
\begin{itemize}
\item an encoding function of the sender, $\mathsf{E}: \mathbb{F}_q^m \rightarrow \mathbb{F}_q^\ell$; and 
\item for each receiver $H\in\mathbb{U}$, a decoding function $\mathsf{G}_H: \mathbb{F}_q^\ell \times \mathbb{F}_q^{|H|} \rightarrow \mathbb{F}_q$, such that $\mathsf{G}_H(\mathsf{E}(X_{[1:m]}),X_H) = X_i$, for some $i \in [1:m]\setminus H$.
\end{itemize}

Define \textit{decoding choice} $D$ as follows:
\begin{equation}
  D: \mathbb{U} \rightarrow [1:m], \text{ such that } D(H) \in [1:m] \setminus H.
\end{equation}
Here, $D(H)$ is the message decoded by receiver $H$. 

The above formulation requires the decoding of only one message at each receiver. Lastly, the aim is to find the optimal broadcast rate for a particular message size $q$, denoted by $\beta_q := \min_{\mathsf{E}, \{\mathsf{G}\}} \ell$ and the optimal broadcast rate over all $q$, denoted by $\beta := \inf_q \beta_q$.

\section{A Motivating Example}\label{sec:motivating-example}
We will now use an example to illustrate two ideas proposed in this paper.
% \subsection{Problem setup} \label{subsec:motivating-example}
Consider a pliable-index-coding problem $\mathcal{P}_1$ with six messages and each receiver requires one new message. All receivers are present except receivers $H_1=\{3\}$, $H_2=\{1,2,3,4\}$, and $H_3=\{3,4,5,6\}$. $\mathcal{P}_1$ does not fall into any category for which the optimal rate $\beta_q(\mathcal{P}_1)$ is known.

 \subsection{Existing lower bounds}
 We have previously established a lower bound~\cite{ongvellambikliewer2019}
\begin{equation}
\beta_q \geq  m - L_\text{max}, \label{eq:previous-lower-bound-longest-chain}
\end{equation}
where $L_\text{max}$ is the maximum length of any nested chain of absent receivers, that is, $H_1 \subsetneq H_2 \subsetneq \cdots \subsetneq H_{L_\text{max}}$, with each $H_i \in \absent$.  In $\mathcal{P}_1$, $L_\text{max} =2$, which can be obtained from $H_1 \subsetneq H_2$ or $H_1 \subsetneq H_3$. So, $\beta_q(\mathcal{P}_1) \geq 6-2=4$.

This lower bound can also be obtained by considering another pliable-index-coding problem $\mathcal{P}_1^-$ formed by removing all receivers each having at least one and up to four messages. % $\mathcal{P}_1^-$ is a complete-$S$ problem with $S = [0:5] \setminus [1:4]$.
It has been shown~\cite{liutuninetti17} that $\beta_q(\mathcal{P}_1^-) = 4$. Combined with the result $\beta_q(\mathcal{P}_1) \geq \beta_q(\mathcal{P}_1^-)$~\cite{ongvellambikliewer2019}, we get $\beta_q(\mathcal{P}_1) \geq 4$.

% \begin{algorithm}[t]
% \caption{\cite{ongvellambikliewer2019} An existing algorithm to construct a decoding chain with skipped messages, which is Algorithm~\ref{algo:2} without Option~2.}
% \label{algo:1}
% \end{algorithm}
\begin{algorithm}[h]
\SetKwInOut{Input}{input}
\SetKwInOut{Output}{output}

\Input{$\mathcal{P}_{m,\mathbb{U},D}$} 
\Output{A \textit{decoding chain} $C$ (a totally ordered set with a total order $\preceq$) and  a set of \textit{skipped messages} $S$}
$C \leftarrow \emptyset$; \texttt{\scriptsize\color{blue} (initialise $C$)}\\
$S \leftarrow \emptyset$; \texttt{\scriptsize\color{blue} (initialise $S$)}\\
\While{$C \neq [1:m]$}{
  \If(\texttt{\scriptsize\color{blue} (receiver $C$ is absent)}){$C \notin \mathbb{U}$}{
    Choose any of the following options:\\
    \SetAlgoVlined
    \SetKwProg{Fn}{Option}{:}{}
    
  \Fn(\texttt{\scriptsize\color{blue} (skip a message)}){\textbf{\upshape 1}}{
      Choose any $a \in [1:m] \setminus C$; \texttt{\scriptsize\color{blue} (skip $a$)}\\
      $C \leftarrow C \cup \{a\}$; \texttt{\scriptsize\color{blue} (expand $C$)}\\
       Define $i \preceq a,$ for all $i \in C$;  \texttt{\scriptsize\color{blue} (define order in $C$)}\\ 
    $S \leftarrow S \cup \{a\}$; \texttt{\scriptsize\color{blue} (expand $S$)}
  }

  \Fn(\texttt{\scriptsize\color{blue} (avoid skipping)}){\textbf{\upshape 2}}{
      Choose any present receiver~$B \subsetneq C$, such that $D(B) \notin C$;\\
    \texttt{\scriptsize\color{blue} (look for a subset $B$,  a present receiver)}\\
    $C \leftarrow C \cup \{D(B)\}$;\\
    \texttt{\scriptsize\color{blue} (add the message that receiver $B$ decodes)}\\
     Define $i \preceq D(B),$ for all $i \in C$;  \\ \texttt{\scriptsize\color{blue} (define order in $C$)}
  }
    
  }
  \Else (\texttt{\scriptsize\color{blue} (receiver $C$ is present)})
  {
    $C \leftarrow C \cup \{D(C) = x\}$;\\ 
    \texttt{\scriptsize\color{blue} (add the message that receiver $C$ decodes)}\\
     Define $i \preceq x,$ for all $i \in C$;  \texttt{\scriptsize\color{blue} (define order in $C$)}
    }
  }
\caption{A new and generalised algorithm to construct a decoding chain with skipped messages}
\label{algo:2}
\end{algorithm}

% \subsection{An existing algorithm to construct a decoding chain with skipped messages}

% \begin{algorithm}[h]
% \SetKwInOut{Input}{input}
% \SetKwInOut{Output}{output}

% \Input{$\mathcal{P}_{m,\mathbb{U},D}$} 
% \Output{A \textit{decoding chain} $C$ (a totally ordered set with a total order $\preceq$) and  a set of \textit{skipped messages} $S$}
% $C \leftarrow \emptyset$; \texttt{\scriptsize\color{blue} (initialise $C$)}\\
% $S \leftarrow \emptyset$; \texttt{\scriptsize\color{blue} (initialise $S$)}\\
% \While{$C \neq [1:m]$}{
%   \If(\texttt{\scriptsize\color{blue} (receiver $C$ is absent)}){$C \notin \mathbb{U}$}{
%     Choose any $a \in [1:m] \setminus C$;\\ \texttt{\scriptsize\color{blue} ($a$ is called a skipped message)}\\
%     $C \leftarrow (C \cup \{a\}$, with $i \preceq a,$ for all $i \in C$);\\ \texttt{\scriptsize\color{blue} (expand $C$)}\\
%     $S \leftarrow S \cup \{a\}$;\\ \texttt{\scriptsize\color{blue} (expand $S$)}
%   }
%   \Else (\texttt{\scriptsize\color{blue} (receiver $C$ is present)})
%   {
%     $C \leftarrow (C \cup \{D(C)\}$, with $i \preceq D(C),$ for all $i \in C$);\\
%     \texttt{\scriptsize\color{blue} (add the message that receiver $C$ decodes)}
%     }
%   }
% \caption{An algorithm to construct a decoding chain with skipped messages}
% \label{algo:1}
% \end{algorithm}
 Another lower bound can be obtained by using our previously proposed algorithm~\cite{ongvellambikliewer2019} to construct a decoding chain of messages. Our previous algorithm is a restriction of our improved Algorithm~\ref{algo:2} devised in this paper, in which  we have defined $\mathcal{P}_{m,\mathbb{U},D}$ as a pliable-index-coding problem with $m$ messages, receivers $\mathbb{U}$, and decoding choice $D$.
If in lines~4--16 of Algorithm~\ref{algo:2}, we always choose Option~1 instead of Option~2, we will retrieve our previous algorithm, which for brevity we will refer to as Algorithm~2 in this paper. Using Algorithm~2 on problem $\mathcal{P}_1$, the following lower bound has been shown~\cite{ongvellambikliewer2019}:
\begin{equation}
  \beta_q(\mathcal{P}_1) \geq m - \max_D |S|, \label{eq:previous-lower-bound}
\end{equation}
where the maximisation is taken over all possible decoding choices $D$ of the receivers, and $S$ is the set of skipped messages obtained from any instance of Algorithm~2 for a specific $D$.

 For Algorithms~\ref{algo:2} and 2, we say that the algorithm ``hits'' a (present or absent) receiver~$C$ if it constructs $C$ upon the execution of lines~8, 14, or 19.
%The algorithm skips a message whenever it ``hits'' a receiver $C$ that is absent (i.e., the \textit{if} condition in the \textit{while} loop is true). 

 For $\mathcal{P}_1$, there exists $D$ for which Algorithm~2 will always hit two absent receivers (either $H_1$ and $H_2$, or $H_1$ and $H_3$) regardless of which messages we skip. This gives a lower bound $\beta_q(\mathcal{P}_1) \geq m - 2 = 4$. To see this, note that receiver~$\emptyset$ is present. Let $D(\emptyset) = 3$. Executing line~19 of the algorithm, we hit $C=\{D(\emptyset)\}=H_1$.  Since receiver~$H_1$ is absent, we execute lines 6--10. Supposing that we skip message~$1$, we will hit $C=\{3,1\}$. Let $D(\{3,1\}) = 2$ and $D(\{3,1,2\}) = 4$. Repeating lines~19--21, we will hit the second absent receiver~$H_2 = \{3,1,2,4\}$. So, by defining $D$ in such a way that no matter which message we choose to skip after hitting $H_1$, the messages to be subsequently added to $C$ stay within $H_2$ or within $H_3$ (until we hit $H_2$ or $H_3$ respectively), we will always hit  $H_2$ or $H_3$.

\subsection{Two new ideas} \label{section:ideas}
We will explain the new ideas in this paper by juxtaposing them with Algorithm~2.
Since skipping fewer messages gives a tighter lower bound, we introduce the following new ideas to skip fewer messages compared to Algorithm~2:
\begin{enumerate}[(a)]
\item\label{idea:1} \textbf{Avoid skipping messages:} This is done by using the subsets of $C$. Using Algorithm~2, when the algorithm hits $C$, and if receiver~$C$ is absent, we skip a message. In our new algorithm, even if receiver~$C$ is absent, if there exists a receiver $B \subsetneq C$ such that $D(B) \notin C$, then the decoding chain can continue by adding $D(B)$ into the chain $C$ without skipping a message.
\item \textbf{Look ahead then skip messages:} Instead of arbitrarily selecting a message~$a \in [1:m] \setminus C$ in Option 1, we will base our choice of skipped messages on $D$. More specifically, we skip a specially chosen message such that the next absent receiver $C$ to be hit will contain a receiver $B \subsetneq C$ whose decoding choice $D(B) \notin C$, and using idea~\ref{idea:1}, we need not skip a message.
\end{enumerate}

\subsection{A new lower bound}
Using the above-mentioned ideas, we now construct a new lower bound for $\mathcal{P}_1$. Note that for any $D$, if fewer than two absent receivers are hit, then $|S| \leq 1$, and this can only lead to the right-hand side of~\eqref{eq:previous-lower-bound} evaluated to 5 or more. So, we only need to consider scenarios where two absent receivers are hit, and in this case the first one must be $H_1$.

To work out the appropriate choice of skipped message upon hitting $H_1$, we \textit{look ahead} and consider $D(H_2 \cap H_3) = D(\{3,4\}) = x$. It is necessary that $x \in H_i \setminus H_j$ for some $i,j \in \{2,3\}$ and $i \neq j$. We then skip any message $y \in H_j \setminus H_i$, and update the decoding chain as $C \leftarrow (C \cup \{y\})$. As $y$ is in $C$ now, the only remaining absent receiver that can be hit is $H_j$. If $H_j$ is not hit, then the algorithm terminates with $|S|=1$; otherwise, it hits $H_j$.

When $H_j$ is hit, we can \textit{avoid} skipping a message by noting that \textsf{(i)}~there is a present receiver~$H_2 \cap H_3 \subsetneq H_j$, and \textsf{(ii)}~it decodes~$D(H_2 \cap H_3) = x \notin H_j$. The decoding chain continues and terminates without hitting another absent receiver.

This means for any $D$, we can always choose $S$ such that $|S| \leq 1$. This gives a lower bound of $\beta_q(\mathcal{P}_1) \geq 6-1=5$. This bounds can be shown to be tight by using a cyclic code for achievability.

More generally, we have the following proposition (which will be proven rigorously later):
\begin{proposition} \label{proposition:3-absent}
  Consider a pliable-index-coding problem $\mathcal{P}_{m, \mathbb{U}}$, where the set of absent receivers is $\absent = \{H_1, H_2, H_3\}$, such that $H_1 \subsetneq H_2 \cap H_3$, and $H_2 \cup H_3 = [1:m]$. We have $\beta_q(\mathcal{P}_{m, \mathbb{U}}) = \beta_q(\mathcal{P}_{m, \mathbb{U}}) = m-1$.
\end{proposition}

\section{A New and Generalised Algorithm} \label{sec:new-algo}

Compared to Algorithm~2, the new Algorithm~\ref{algo:2} has Option 2, which implements the two new ideas in Section~\ref{section:ideas}.
It allows us to avoid skipping a message even when an absent receiver $C$ is hit, as long as a suitable present receiver $B \subsetneq C$ can be found. If Option~1 is always selected, we revert back to Algorithm~2 as a special case. Although choosing Option~1 may seem counter-intuitive, we will see that later that choosing Option~1 simplifies the proof of our results as it avoids evaluating $D(B)$ required in Option~2.

The sketch of proof for the lower bound~\eqref{eq:previous-lower-bound} for Algorithm~2 is as follows~\cite{ongvellambikliewer2019}:  We started with a bipartite graph $G_D$ that describes $\mathcal{P}_{m,\mathbb{U},D}$. We showed that for each instance of Algorithm~2, there is a series of pruning operations on $G_D$ that yield an acyclic graph $G_D'$ with $m-|S|$ remaining messages. The graph $G_D$ is acyclic because, by construction, all directed edges flow from message nodes that are \textit{larger} to message nodes that are \textit{smaller} with respect to the order $\preceq$.  As $m-|S|$ is a lower bound on $\mathcal{P}_{m,\mathbb{U},D}$~\cite[Lem.~1]{neelytehranizhang13}, and that $\beta_q(\mathcal{P}_{m,\mathbb{U}}) = \min_D \beta_q(\mathcal{P}_{m,\mathbb{U},D})$, we have \eqref{eq:previous-lower-bound}.

 We now show that the lower bound \eqref{eq:previous-lower-bound} is still valid using Algorithm~\ref{algo:2}.
 Algorithm~\ref{algo:2} differs from Algorithm~2 by having Option~2. Using Option~2 on a present receiver~$B$, this receiver is preserved (that is, not removed during the pruning operation) in the graph $G_D$. With this additional receiver not removed (compared to Algorithm~2), there are additional directed edges flowing from the a \textit{larger} message node to \textit{smaller} message nodes with respect to the order $\preceq$, that is, from the message node $D(B)$ to message nodes $\{x \in B\}$ through the receiver node $B$. Clearly, all additional edges retained due to Option~2 in Algorithm~\ref{algo:2} do not create any directed cycle. Hence, the proof for the lower bound~\eqref{eq:previous-lower-bound} for Algorithm~2 can be modified accordingly to give the following:

 \begin{lemma} \label{lemma:chain-lower-bound-new} 
  Consider a pliable-index-coding problem $\mathcal{P}_{m,\mathbb{U}}$. For a specific $D$, let $S$ be the set of skipped messages for an instance of Algorithm~\ref{algo:2}. Then,
  \begin{equation}
    \beta_q (\mathcal{P}_{m,\mathbb{U}}) \geq  m - \max_D  |S|. \label{eq:chain-lower-bound-new}
  \end{equation}
\end{lemma}
The lower bound is obtained by maximising $|S|$ over all decoding choices $D$. By optimising the choice of skipped messages for each $D$ such that the minimum number of messages is skipped, we obtain the following lower bound:
\begin{equation}
    \beta_q (\mathcal{P}_{m,\mathbb{U}}) \geq  m - \max_D \min_{S}  |S| = m - L^*, \label{eq:chain-lower-bound-new-2}
  \end{equation}
  where we define
  \vspace*{-1.8ex}
\begin{equation}
  L^* := \max_D \min_S |S|. \label{eq:chain-lower-bound}
\end{equation}

 % With this, the lower bound~\eqref{eq:chain-lower-bound-new-2} can also be written as
% \begin{equation}
% \beta_q (\mathcal{P}_{m,\mathbb{U}}) \geq  m - L^*. \label{eq:lower-bound-L-star}
% \end{equation}

\begin{remark}
  For any given $D$, although any instance of Algorithm~\ref{algo:2} gives a lower bound for $\beta_q(\mathcal{P}_{m,\mathbb{U},D})$, skipping as few messages as possible gives tighter lower bounds.
\end{remark}

Intuitively, Algorithm~\ref{algo:2} says that the construction of decoding chain $C$ can continue even if receiver $C$ is absent, because if receiver $B \subsetneq C$ can decode $D(B) \notin C$, then knowing $C$, one is able to obtain $D(B)$ to extend the decoding chain.

Before formally deriving the second idea of ``look ahead and skip'' in Section~\ref{sec:look-ahead}, in the next section,  we first improve upon an existing lower bound that can be obtained by simply looking at how the absent receivers are nested, that is, without needing an algorithm that constructs decoding chains.

\section{An Improved Nested-Chain Lower Bound} \label{sec:new-nested-chain}

 From \eqref{eq:chain-lower-bound-new-2}, we see that any upper bound on $L^*$ provides a lower bound on $\beta_q$. For instance, see lower bound~\eqref{eq:previous-lower-bound-longest-chain}, where $L^* \leq L_\text{max}$. 
The lower bound based on $L_\text{max}$ may be loose, because we may be able to skip certain messages to avoid hitting some absent receivers in the longest chain. In this paper, we will prove a better\footnote{The new lower bound is strictly better for certain problems.} lower bound based on this idea. We now prove the following theorem:
\begin{theorem} \label{theorem:improved-nested-chain}
  $L^* \leq L-1$ if the following condition holds:
%   \begin{quote}
%   Condition: For every chain of absent receivers $H_1\subsetneq H_2 \subsetneq \dotsm \subsetneq H_{L'}$ (for some $L' \geq L$), there exist $H_k \cup \{a\}$ (for some $k \in [1:L-1]$ and for some $a \notin H_k$) that is not contained in any nested chain of $L-k$ absent receivers.
% \end{quote}
% or this one
% \begin{quote}
  For every chain of absent receivers of length at least $L$, say, $H_1 \subsetneq \dotsm \subsetneq H_{L'}$ for some $L' \geq L$, where $H_i \in \absent$, there exists $H_k \cup \{a\}$ (for some $k \in [1:L-1]$ and for some $a \notin H_k$) such that there is no chain of absent receivers of length $L-k$ where  $\displaystyle (H_k \cup \{a\}) \subseteq \underbracket{H'_1  \subsetneq \dotsm \subsetneq H'_{L-k}}_{\text{absent-receiver chain}}$, with $H_i' \in \absent$.
%  \end{quote}
\end{theorem}
\vspace{1ex}

\begin{IEEEproof}[Proof of Theorem~\ref{theorem:improved-nested-chain}]
  Recall that each instance of Algorithm~\ref{algo:2} (or Algorithm~2) returns a decoding chain $C = \{c_1, c_2, \dotsc, c_m\}$, in the order $c_i \preceq c_j$ iff $i \leq j$, and a set of skipped messages $S \subseteq C$.

  Let $c_i$ by the $k$th skipped message. This means the algorithm must have hit an absent receiver~$H\in \absent$, where
  \begin{equation}
    H = \begin{cases}
      \emptyset, &\text{if }  i=1, \\
      \{c_1, \dotsc, c_{i-1}\}, & \text{otherwise } (\text{i.e., }i \in [2:m]).
    \end{cases}
  \end{equation}

 Suppose that $\ell$ is the maximum number of absent receivers that can form a chain $(H \cup \{c_1\}) \subseteq H'_1 \subsetneq H'_2 \subsetneq \cdots \subsetneq H'_\ell$, with each $H'_i \in \absent$. Then, at most $\ell$ more absent receivers can be hit. Consequently, the algorithm must terminate with $|S| \leq k + \ell$.

  Now, for all nested receiver chains of length $L$ or larger, suppose that the condition stated in the theorem is true, we can always skip receiver $a$ after hitting  $H_k$, such that $|S| < k + (L - k)$. As $|S|$ is an integer, $|S| \leq L-1$. Since this is true for all nested receiver chains of length $L$ or larger, we can always avoid skipping $L$ messages, giving $L^* \leq L-1$. 
\end{IEEEproof}

We will show the utility of Theorem~\ref{theorem:improved-nested-chain} using an example:
\begin{figure}[t]
  \centering
  \includegraphics[scale=0.5]{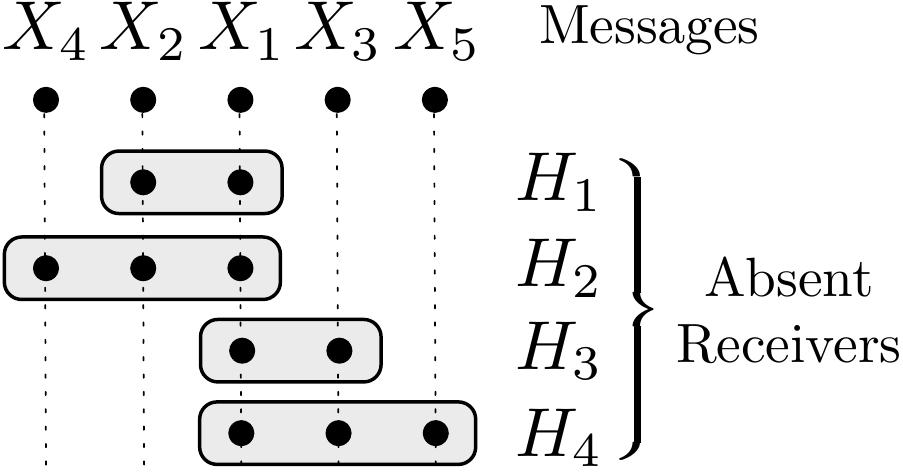}
  \caption{Pliable-index-coding problem $\mathcal{P}_2$ for Example~\ref{ex2}}
  \label{fig:ex2}
\end{figure}

\begin{example}\label{ex2}
  Consider $\mathcal{P}_2$ with five messages and four absent receivers $H_1=\{1,2\}, H_2=\{1,2,4\}, H_3=\{1,3\}$, and $H_4=\{1,3,5\}$, as depicted in Figure~\ref{fig:ex2}. The length of the longest nested absent-receiver chain is 2. Our previous lower bound gives $\beta_q \geq 3$ (see~\eqref{eq:previous-lower-bound-longest-chain}). Now, we invoke Theorem~\ref{theorem:improved-nested-chain}, and consider all chains of length $L\geq 2$, which are $H_1 \subsetneq H_2$ and $H_3 \subsetneq H_4$.
  \begin{itemize}
  \item When $H_1$ is hit, we skip message~3. $\{1,2,3\}$ is not contained in any absent receiver.
  \item When $H_3$ is hit, we skip message~4. $\{1,3,4\}$ is not contained in any absent receiver.
  \end{itemize}
So, we have $L^* \leq 1$. Noting \eqref{eq:chain-lower-bound} and \eqref{eq:chain-lower-bound-new-2}, we get $\beta_q \geq 5-1=4$. This lower bound can be achieved by the code~$(X_3+X_5,\, X_1,\, X_2,\, X_4)$.
\end{example}

While this new nested-chain lower bound improved on our previous longest-chain lower bound, it is still insufficient to solve $\mathcal{P}_1$ described in Section~\ref{sec:motivating-example}. To solve $\mathcal{P}_1$, we will use the  ``look ahead and skip'' technique detailed in the next section.

\section{Skipping Messages with Look Ahead} \label{sec:look-ahead}

In this section, when we hit an absent receiver, say $H \subseteq [1:m]$,  we will propose a method to skip a message in such a way to guarantee that we will subsequently not need to skip any message when we hit any absent receiver from a special subset of absent receivers. Let this subset of absent receivers be $\mathbb{A} \subseteq \absent \setminus \{H\}$. By definition, all absent receivers in $\mathbb{A}$ are supersets of $H$. This method is used in conjunction with Algorithm~\ref{algo:2}.
We will prove the following result:
\begin{theorem}\label{theorem:look-ahead-skip}
  Let $H \in \absent$ by an absent receiver, and $\mathbb{A} \subseteq \absent \setminus \{H\}$ be a subset of absent receivers that belongs to any of the following cases, where $H \subsetneq H'$ for all $H' \in \mathbb{A}$.
  Running Algorithms~\ref{algo:2}, suppose that $H$ is hit. We can always choose to skip a message such that, if any $H' \in \mathbb{A}$ is hit subsequently, we can avoid skipping a message.
\begin{enumerate}
\item\label{case1} $\mathop{\bigcup}_{H' \in \mathbb{A}} H' \neq [1:m]$.
%\item[]\hspace*{-4ex} For the rest of the cases, $\mathop{\bigcup}\limits_{S' \in \mathbb{S}} S' = [1:m]$.
\item\label{case2} $\mathbb{A}$ is a minimal cover\footnote{A family of sets $\mathbb A = \{A_\ell: \ell \in L\}$ is a minimal cover of $B$ iff $\mathop{\bigcup}_{\ell\in L} A_\ell = B$, and for any strict subset $L’ \subsetneq L$, $\mathop{\bigcup}_{\ell \in L’} A_\ell \subsetneq B.$} of $[1:m]$, $T :=\mathop{\bigcap}_{H' \in \mathbb{A}} H' \supsetneq H$, and $T \in \mathbb{U}$.
\item \label{case3} $\mathbb{A}$ is a minimal cover of $[1:m]$, and $\mathop{\bigcap}_{H' \in \mathbb{A}} H'= H$; furthermore, there exist\footnote{If this is false, $\mathbb{A} \cup \{H\}$ forms 1-truncated $L$-nested absent receivers, which we will define in Definition~\ref{def:truncated} later.} $H_1, H_2 \in \mathbb{A}$ such that $T:= H_1 \cap H_2 \supsetneq H$ and $T \in \mathbb{U}$.
\end{enumerate}
\end{theorem}

\begin{IEEEproof}[Proof of Theorem~\ref{theorem:look-ahead-skip}]
  For case~\ref{case1}, by skipping any $a \in [1:m] \setminus \left( \mathop{\bigcup}_{H' \in \mathbb{A}} H' \right)$, we will not hit any absent receiver in $\mathbb{A}$.

  For case~\ref{case2}, we \textit{look ahead} and check $D(T)$. Since receiver~$T$ is present, $D(T)$ is defined. As $T := \mathop{\bigcap}_{H' \in \mathbb{A}} H'$ and $D(T) \notin \mathop{\bigcap}_{H' \in \mathbb{A}} H'$,  there must exist an absent receiver~$H_1 \in \mathbb{A}$ that does not contain $D(T)$. As $\mathbb{A}$ is a minimal cover, there exists some $a \in H_1$ that is not in all other sets in $\mathbb{A}$, that is, $a \notin \mathop{\bigcup}_{H' \in \mathbb{A} \setminus \{H_1\}} H'$. We choose to skip $a$, and by doing so, we will never hit any receiver in $\mathbb{A} \setminus \{H_1\}$. If we hit $H_1$, we can choose Option~2 in the algorithm without needing to skip any message, since $T \subseteq H_1$ and $D(T) \notin H_1$.

  For case~\ref{case3}, we \textit{look ahead} and check $D(T)$. As receiver~$T$ is present, $D(T)$ is defined. $D(T) \notin T = H_1 \cap H_2$. Without loss of generality, suppose $D(T) \notin H_1$. When we follow the same argument for case~\ref{case2} by skipping some $a \in H_1$ that is not in all other sets in $\mathbb{A}$. By doing so, will can always avoid skipping a message due to hitting $H_1$.
\end{IEEEproof}

\section{Applications of Results} \label{sec:application}

\subsection{Optimal rates for the slightly imperfect $L$-nested setting}

We have previously defined a class of pliable-index-coding problems as follows~\cite{ongvellambikliewer2019}:
\begin{definition}
  A pliable-index-coding problem is said to have \textit{perfect $L$-nested absent receivers} iff the messages $[1:m]$ can be partitioned into $L+1 \in [2:m]$ subsets $P_0, P_1, \dotsc, P_{L}$ (that is, $\mathop{\bigcup}_{i=0}^L P_i = [1:m]$ and $P_i \cap P_j = \emptyset$ for all $i \neq j$), such that only $P_0$ can be an empty set, and there are exactly $2^L-1$ \textit{absent} receivers, which are defined as
  \begin{equation}
 \textstyle    H_Q := P_0 \cup \left( \mathop{\bigcup}_{i \in Q} P_i \right), \text{ for each } Q \subsetneq [1:L].
  \end{equation}
\end{definition}

% \begin{figure}[t]
%   \centering
%   \includegraphics[scale=0.9]{43}
%   \caption{Perfect $3$-nested absent receivers}
%   \label{fig:perfect}
% \end{figure}
% Figure~\ref{fig:perfect} depicts an example of perfect $3$-nested absent receivers.
For any pliable-index-coding problem $\mathcal{P}_{m,\mathbb{U}}$ with perfect $L$-nested absent receivers, $\beta_q(\mathcal{P}_{m,\mathbb{U}}) = m-L$~\cite{ongvellambikliewer2019}.

\begin{figure}[t]
  \centering
  \includegraphics[scale=0.5]{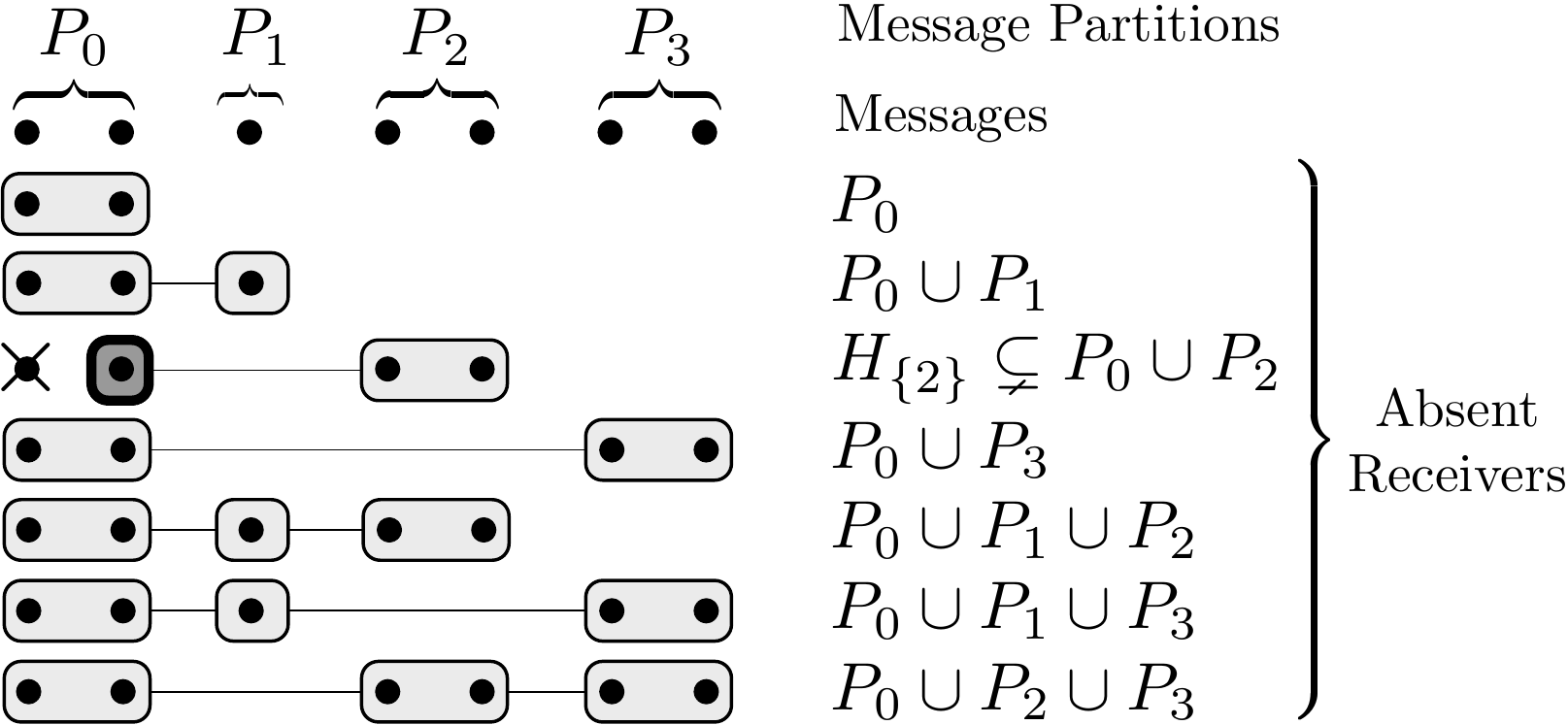}
  \caption{Slightly imperfect $3$-nested absent receivers, formed by shrinking the side-information set of one receiver among perfect $3$-nested absent receivers.}
  \label{fig:imperfect}
\end{figure}

With Theorem~\ref{theorem:look-ahead-skip}, we can determine the optimal rate of problems deviating from the perfect $L$-nested setting.
We now prove the optimal rate for pliable-index-coding problems with slightly imperfect $L$-nested absent receivers. Figure~\ref{fig:imperfect} depicts a example of slightly imperfect $3$-nested absent receivers.

\begin{theorem}\label{theorem:less-perfect}
  Consider a pliable-index-coding problem $\mathcal{P}_{m,\mathbb{U}}$ that comprises perfect $L$-nested absent receivers with the following change: one absent receiver $H_Q = P_0 \cup \left( \mathop{\bigcup}_{i \in Q} P_i \right)$, for some $Q \subsetneq [1:L]$, is changed to the  absent receiver $H_Q \subsetneq P_0 \cup \left( \mathop{\bigcup}_{i \in Q} P_i \right)$.
 %  \begin{equation}
 %   \textstyle  \tilde{H}_Q \ \subsetneq P_0 \cup \tilde{P}_j \cup \left( \mathop{\bigcup}\limits_{i \in Q \setminus \{j\}} P_i \right),
 % \end{equation}
 % for some $j \in Q$ and $\tilde{P}_j \subsetneq P_j$.
 Then, $\beta_q(\mathcal{P}_{m,\mathbb{U}})  = m - L + 1$.
\end{theorem}

\indent\indent \textit{Proof of 
Thm~\ref{theorem:less-perfect}:}  See 
\ifx\arxiv\undefined
 the extended version of this paper~\cite{ongvellambikliewersadeghi-arvix-2019}.
\else
Appendix.
\fi

%With Theorem~\ref{theorem:less-perfect},
We can now prove 
Proposition~\ref{proposition:3-absent} that we stated earlier.
\begin{IEEEproof}[Proof of Proposition~\ref{proposition:3-absent}]
  $\mathcal{P}_{m,\mathbb{U}}$ is formed by having perfect 2-nested absent receivers with $P_0 = H_2 \cap H_3$, $P_1 = H_2 \setminus H_3$, $P_2 = H_3 \setminus H_2$, and then replacing absent receiver $P_0$ with $H_1 \subsetneq P_0$. Using Theorem~\ref{theorem:less-perfect}, we have $\beta_q(\mathcal{P}_{m,\mathbb{U}})  = m - 1$.
%  Let $S= H_1$ and $\mathbb{S} = \{ H_2, H_3\}$. Invoking Theorem~\ref{theorem:look-ahead-skip} (Case~\ref{case2}), the maximum number of skipped messages $L^* \leq 1$, and substituting it into \eqref{eq:lower-bound-L-star} gives $\beta_q(\mathcal{P}_{m,\mathbb{U}}) \geq m-1$. The lower bound can be achieved by sending $X_{H_2}$ uncoded and the rest using a cyclic code.
\end{IEEEproof}

\subsection{Optimal rates for $T$-truncated  $L$-nested absent receivers}

We define another variation of perfect $L$-nested absent receivers.
\begin{definition} \label{def:truncated}
  A pliable-index-coding problem is said to have \textit{$T$-truncated $L$-nested absent receivers} iff the messages $[1:m]$ can be partitioned into $L+1 \in [2:m]$ subsets $P_0, P_1, \dotsc, P_{L}$ (that is, $\mathop{\bigcup}_{i=0}^L P_i = [1:m]$ and $P_i \cap P_j = \emptyset$ for all $i \neq j$), such that only $P_0$ can be an empty set, and there are $\sum_{i=0}^T \binom{L}{i}$ absent receivers,  which are defined as 
  \begin{equation}
 \textstyle    H_Q = P_0 \cup \left( \mathop{\bigcup}_{i \in Q} P_i \right),\;\; \forall Q \subsetneq [1:L],\text{ with } |Q| \in [0:T],
\end{equation}
for some $T \in [0:L-1]$.
\end{definition}

\begin{figure}[t]
  \centering
  \includegraphics[scale=0.5]{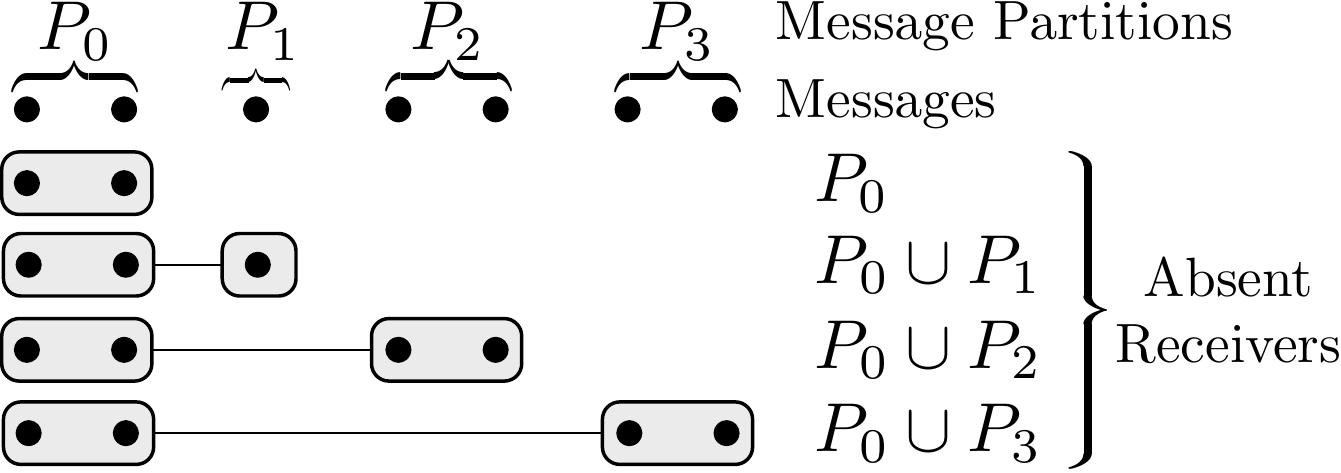}
  \caption{$1$-truncated $3$-nested absent receivers, formed by keeping the top few groups of perfect $3$-nested absent receivers}
  \label{fig:truncated}
\end{figure}
Note that $(L-1)$-truncated $L$-nested absent receivers are equivalent to perfect $L$-nested absent receivers. Figure~\ref{fig:truncated} depicts an example of $1$-truncated $3$-nested absent receivers.

\begin{theorem} \label{theorem:truncated-nested}
  For any pliable-index-coding problem $\mathcal{P}$ with $T$-truncated $L$-nested absent receivers, $\beta(\mathcal{P}) = \beta_q(\mathcal{P}) = m-T-1$, for sufficiently large $q$.
\end{theorem}

\indent\indent \textit{Proof of 
Thm~\ref{theorem:truncated-nested}:}  See
\ifx\arxiv\undefined
 the extended version of this paper~\cite{ongvellambikliewersadeghi-arvix-2019}.
\else
Appendix.
\fi

\subsection{Optimal rates for a small number of absent receivers}

We have established that $\beta_q = m$ if and only if there is no absent receiver, that is $|\absent|=0$.

\begin{corollary}
  If $1 \leq |\absent| \leq 2$, then $\beta_q=m-1$.
\end{corollary}
\begin{IEEEproof}
  For $|\absent|=1$, by definition, the absent receiver $H \subsetneq [1:m]$, and hence $\mathop{\bigcup}_{H \in \absent} H \neq [1:m]$. So, the result follows from~\cite[Thm.~1]{ongvellambikliewer2019}.
  For $|\absent|=2$, there can be either no nested pair or one nested pair of absent receivers. The result follows from~\cite[Thm.~3]{ongvellambikliewer2019}.
\end{IEEEproof}

While the optimal rate for up to two absent receivers can be determined using our previous results, we need the new results presented in this paper for more absent receivers.

\begin{theorem} \label{theorem:3-absent}
  Suppose $|\absent|=3$. Then
  \begin{equation*}
    \beta_q =
    \begin{cases}
      m-2, & \text{ if the absent receivers are perfect 2-nested},\\
      m-1, & \text{ otherwise}.
    \end{cases}
  \end{equation*}
\end{theorem}

\begin{theorem}\label{theorem:4-absent}
  Suppose $|\absent|=4$. Then
  \begin{equation*}
    \beta_q =
    \begin{cases}
      m-2, & \text{ if a subset of absent receivers is either}\\ & \text{ perfect 2-nested or 1-truncated 3-nested},\\
      m-1, & \text{ otherwise}.
    \end{cases}
  \end{equation*}
\end{theorem}

\indent\indent \textit{Proofs of Thms~\ref{theorem:3-absent} and \ref{theorem:4-absent}:}  See
\ifx\arxiv\undefined
 the extended version~\cite{ongvellambikliewersadeghi-arvix-2019}.
\else
Appendix.
\fi

%For the cases where $|\absent| \leq 4$, we see that only structured case results in savings beyond one coded symbol.

%\section{Conclusions}

%\vspace*{-1.5ex}

\bibliography{../bib}

\ifx\arxiv\undefined
\else

\appendix

\begin{IEEEproof}[Proof of Theorem~\ref{theorem:less-perfect}]
  We will use Algorithm~\ref{algo:2} and show that $L^*  \leq L-1$. Note that the length of any longest chain of nested absent receivers in $\mathcal{P}_{m,\mathbb{U}}$ is $L$. %, and any of these chain must be in the form $H_\emptyset \subsetneq H_{A_1} \subsetneq H_{A_2} \subsetneq \dotsc \subsetneq H_{A_{L-1}}$, for some distinct $\emptyset \subsetneq A_1 \subsetneq \dotsm \subsetneq A_{L-1} \subsetneq [1:L]$, where $|A_i|=i$.
  
  We will now consider all decoding choices $D$ that can potentially result in $L$ absent receivers being hit. We need not consider any other decoding choices, as we will hit at most $L-1$ absent receivers and skip at most $L-1$ messages. Without loss of generality, let $Q = \{1, 2, \dotsc, |Q|\}$.

  If $|Q|=0$, then the first receiver to be hit must be $H_\emptyset \subsetneq P_0$ (otherwise, we will not hit at most $L-1$ absent receivers in total). We invoke Theorem~\ref{theorem:look-ahead-skip} (Case 2), where
  $H = H_\emptyset$, $\mathbb{H} =  \Big\{ (P_0  \cup  P_\ell): \ell \in [1:L] \Big\}$, and  $T = P_0$. To hit $L$ absent receivers in total, the next absent receiver to be hit must be from $\mathbb{H}$. From Theorem~\ref{theorem:look-ahead-skip}, we know that we need not skip any message when we hit any absent receiver in $\mathbb{H}$. So, in total, we will skip at most $L-1$ messages.

  Otherwise, for $|Q| \in [1:L-1]$, we will first hit $P_0$. Then, we perform and repeat the following step
  \begin{itemize}
  \item When we hit $(P_0 \cup \dotsc \cup P_t)$, for $0 \leq t \leq |Q|-1$,  we skip some
    \begin{equation}
      a \in
      \begin{cases}
        P_{t+1} \setminus H_Q, &\text{if } H_Q \cap P_{t+1} \neq P_{t+1},\\
        P_{t+1}, & \text{otherwise.}
      \end{cases}
    \end{equation}

  \end{itemize}
  By doing so, if $0 \leq t \leq |Q|-2$, we will next hit $(P_0 \cup \dotsc \cup P_{t+1})$. In the last step when $t = |Q|-1$, we observe the following:
  \begin{enumerate}
  \item If $H_Q \cap P_{|Q|} \neq P_{|Q|}$, we skip any $a \in P_{|Q|} \setminus H_Q$. With this choice of skipped message, we will not hit $H_Q$.
  \item Otherwise, we have $H_Q \cap P_j \neq P_j$ for some $j \in [0:|Q|-1]$. We skip any $a \in P_{|Q|}$. Since the decoding chain already contains  $(P_0 \cup \dotsm \cup P_{|Q|-1})$, we will also not hit $H_Q$.
  \end{enumerate}
Consequently, the next receiver to be hit can only be $H_{Q \cup \{q\}}$, for some $q \in \{|Q|+1, \dotsc, L \}$. So, the total absent receivers hit is at most $L-1$.

 We have shown that regardless of which decoding choice, we will skip at most $L-1$ messages, and $L^* \leq L-1$.

  We showed that sending $X_{P_0}$ uncoded and $X_{P_i}$ using cyclic codes for each $i \in [1:L]$ achieves $m-L$ for perfect $L$-nested absent receivers~\cite{ongvellambikliewer2019}. In comparison, this problem $P_{m, \mathbb{U}}$ contains an additional present receiver $P_0 \cup \left( \mathop{\bigcup}\limits_{i \in Q} P_i \right)$. To satisfy this receiver, we transmit another message $X_a$ for some $a \in P_k$ for some $k \in [1:L] \setminus Q$. This codelength for this code is thus $m-L+1$.  
\end{IEEEproof}

{\ }

\begin{IEEEproof}[Proof of Theorem~\ref{theorem:truncated-nested}]
  It is easy to see that the the longest nested chain in this case is $T+1$, and the chain consists of absent receivers $H_{Q_0} \subsetneq H_{Q_1} \subsetneq \dotsm \subsetneq H_{Q_T}$, where $Q_0 \subsetneq Q_1 \subsetneq \dotsm \subsetneq Q_T$ and $|Q_i|=i$ for all $i \in [0:T]$. So, \eqref{eq:previous-lower-bound-longest-chain} gives a lower bound $\beta_q(\mathcal{P}) \geq m-T-1$.

  Let the problem with  perfect $L$-nested absent receivers on the same partitions $\{P_i: i \in [0:L]\}$ be $\mathcal{P}^-$, where $\beta_q(\mathcal{P}) = m-L$~\cite[Thm.~4]{ongvellambikliewer2019}. Achievability for $\mathcal{P}^-$ is attained by sending messages in $P_0$ uncoded, $X_{P_0} = Y_0 \in \mathbb{F}_q^{|P_0|}$, and messages in each $P_i$, $i \in [1:L]$, using a cyclic code $Y_i = (Z_{i,1}+Z_{i,2},\, Z_{i,2}+Z_{i,3},\, \dotsc,\, Z_{i,|P_i|-1}+Z_{i,|P_i|}) \in \mathbb{F}_q^{|P_i|-1}$, where $Z_{i,j}$ is the $j$th message in $P_i$. Let this code be $Y=(Y_0, \dotsc, Y_L) \in \mathbb{F}_q^{m-L}$.

  We will now add receivers group by group until we get $\mathcal{P}$. At each stage, we compose additional coded messages to satisfy newly added receivers.
  \begin{itemize}
  \item First, we add receivers $H_Q$ with $|Q|=L-1$ to $\mathcal{P}^-$, we will add another coded message to satisfy these added receivers (the other receivers are can decode with $Y$ and their side information). We add $V_{L-1} = \sum_{i=1}^L Z_{i,1} \in \mathbb{F}_q$. Each added receiver knows all but one message in $\{Z_{i,1}: i \in [1:L]\}$, and can then decode a new message from $V_{L-1}$.
  \item Then, we further add receivers $H_Q$ with $|Q|=L-2$. For this, we will further add another coded message to satisfied the newly added receivers. The added coded message is $V_{L-2} = \sum_{i=1}^L \gamma^{i} Z_{i,1} \in \mathbb{F}_q$, where $\gamma$ is a primitive element in $\mathbb{F}_q$. Note each newly added receivers knows all but two messages in $\{Z_{i,1}: i \in [1:L]\}$ and can then decode a new message from $(V_{L-1},V_{L-2}) \in \mathbb{F}_q^2$.
  \item This step is repeated. That means when we add receivers $H_Q$ with $|Q| = L- k$, for $k \in [1:L-1-T]$. We add a coded message $V_{L-k} = \sum_{i=1}^L (\gamma^{k-1})^{i} Z_{i,1} \in \mathbb{F}_q$. Each newly added receiver knows $L-k$ messages in $\{Z_{i,1}: i \in [1:L]\}$, and can then decode a new message from $(V_{L-1}, V_{L-2}, \dotsc, V_{L-k}) \in \mathbb{F}_q^k$ if $q$ is sufficiently large.
  \end{itemize}
So, by sending $(Y_0, Y_1, \dotsc, Y_L, Z_{L-1}, Z_{L-2}, \dotsc, Z_{L-(L-1-T)}) \in \mathbb{F}_q^{m-T-1}$, every receiver in $\mathcal{P}$ can obtain at least one new message. So, the rate of $m-T-1$ is achievable for sufficiently large $q$.  
\end{IEEEproof}

{\ }

\begin{IEEEproof}[Proof of Theorem~\ref{theorem:3-absent}]
  Let the absent receivers be $H_1$, $H_2$, and $H_3$, where the labelling is arbitrary. If $\mathop{\bigcup}\limits_{i=1}^3 H_i \neq [1:m]$, then $\beta_q = m-1$~\cite[Thm.~1]{ongvellambikliewer2019}.

  For the rest of the settings, we have $\mathop{\bigcup}\limits_{i=1}^3 H_i = [1:m]$. For this case, the length of the longest nested chain of absent receivers is at most two. Therefore, there can be at most two pairs of nested absent receivers.
  \begin{itemize}
  \item If there is one or no nested pair of absent receivers, we have
    $\beta_q = m-1$~\cite[Thm.~3]{ongvellambikliewer2019}. 
  \item Otherwise,
    we have two nested absent receiver pairs, and they must have the
    configuration $H_1 \subseteq (H_2 \cap H_3)$, and
    $H_2 \cup H_3 = [1:m]$. For this case, we have two scenarios:
    \begin{itemize}
    \item If $H_1 \subsetneq (H_2 \cap H_3)$,
    Proposition~\ref{proposition:3-absent} gives $\beta_q =
    m-1$. 
  \item Otherwise, $H_1 = H_2 \cap H_3$, which is perfect 2-nested,
    and $\beta_q = m-2$~\cite[Thm.~4]{ongvellambikliewer2019}.
  \end{itemize}

\end{itemize}
The proof is complete by noting that the only the last case is the only case with perfect 2-nested absent receivers.
\end{IEEEproof}

{\ }

\begin{IEEEproof}[Proof of Theorem~\ref{theorem:4-absent}]
  Let the absent receivers be $\{H_i: i \in [1:4]\}$, where the labelling is arbitrary. Again, if $\mathop{\bigcup}\limits_{i=1}^4 H_i \neq [1:m]$, then $\beta_q = m-1$~\cite[Thm.~1]{ongvellambikliewer2019}.
For the rest of the settings, we have $\mathop{\bigcup}\limits_{i=1}^4 H_i = [1:m]$. %For this case, the length of the longest nested chain of absent receivers is $L_\text{max} \leq 3$.

% Suppose that $L_\text{max} = 3$. Then, we must have the configuration $H_1 \subsetneq H_2 \subsetneq H_3$ and $H_3 \cup H_4 = [1:m]$, meaning that $(H_3,H_4)$ is not a nested pair. Now, we invoke Theorem~\ref{theorem:improved-nested-chain}, and consider $L=3$ and the chain $H_1 \subsetneq H_2 \subsetneq H_3$. For $k=1$ note that $H_1 \cup \{a\}$, for some $a \notin H_2$, can only be contained in $H_3$ and $H_4$, which are not nested. So, $H_1 \cup \{a\}$ is not contained in any nested chain of $L-k=2$ absent receivers. Theorem~\ref{theorem:improved-nested-chain} gives $L^* \leq L-1=2$. If $L_\text{max} \leq 2$, $L^* \leq 2$ follows immediately.  So, $\beta_q \geq m-2$ for all cases.

  If the minimum (absent-receiver) cover of $[1:m]$ is four, then there is no nested pair of absent receiver, and $\beta_q = m-1$~\cite[Thm.~3]{ongvellambikliewer2019}.

  If the minimum cover of $[1:m]$ is three, say $H_2 \cup H_3 \cup H_4 = [1:m]$, then $L_\text{max} \leq 2$.
  \begin{itemize}
  \item If $L_\text{max} =1$, then
    \eqref{eq:previous-lower-bound-longest-chain} gives
    $\beta_q \geq m-1$, which is achievable by sending $X_{H_2}$
    uncoded and the rest using a cyclic code. 
  \item Otherwise,
    $L_\text{max}=2$, and so $\beta_q \geq m-2$.
     \begin{itemize}
    \item If there is only one or no
    nested pair,
    $\beta_q = m-1$~\cite[Thm.~3]{ongvellambikliewer2019}. 
  \item If there
    are two nested pairs, say $H_1 \subsetneq H_2 \cap H_3$, 
    then the
    only way to hit two absent receivers is to first hit $H_1$, and
    then hit either $H_2$ or
    $H_3$. Invoking Theorem~\ref{theorem:look-ahead-skip} (case~\ref{case1}) with $H=H_1$
    and $\mathbb{H} = \{H_2, H_3\}$ where $H_2 \cup H_3 \neq [1:m]$, we can always avoid skipping any
    more message after hitting $H_1$. This gives $L^* \leq 1$ and $\beta_q \geq m-1$, which is achievable.
  \item Otherwise, there are three nested pairs $H_1 \subseteq H_1 \cap H_2 \cap H_3$. Let $\mathbb{S} = \{H_2, H_3, H_4 \}$.
    \begin{itemize}
    \item If $H_1 \subsetneq H_2 \cap H_3 \cap H_4$, invoking Theorem~\ref{theorem:look-ahead-skip} (case~\ref{case2}), we can again show that $\beta_q \geq m-1$, which is also achievable.
    \item Otherwise, $H_1= H_2 \cap H_3 \cap H_4$.
      \begin{itemize}
      \item If $H_i \cap H_j = H_1$ for all distinct $i,j \in [2:4]$, meaning that they are 1-truncated 3-nested. Using Theorem~\ref{theorem:truncated-nested}, we get $\beta_q =m-2$. (Note that since $L-1-T=1$, binary codes $q=1$ suffices).
      \item Otherwise, $H_i \cap H_j \supsetneq H_1$ for some distinct $i,j \in [2:4]$. Invoking Theorem~\ref{theorem:look-ahead-skip} (case~\ref{case3}), we can gain show that $\beta_q \geq m-1$, which is also achievable.
      \end{itemize}
 
  \end{itemize}

  \end{itemize}

\end{itemize}

If the minimum cover of $[1:m]$ is two, say $H_3 \cup H_4 = [1:m]$.
\begin{itemize}
  \item If $L_\text{max} =1$, then similar to the argument above, $\beta_q =m-1$.
  \item If $L_\text{max}=2$, then $\beta_q \geq m-2$.
    \begin{itemize}
\item If $H_i = H_3 \cap H_4$ for some distinct $i \in [1:2]$,
    then $(H_i, H_3, H_4)$ is perfect 2-nested. Since this problem
    $\mathcal{P}$ has one additional absent receiver (which is $H_j$,
    $j \neq i$) compared to a problem $\mathcal{P^+}$ with
    three absent (perfect nested) receivers $(H_i, H_3, H_4)$,
    $\beta_q(\mathcal{P}) \leq \beta_q(\mathcal{P}^+) = m-2$. Combining with $\beta_q(\mathcal{P}) \geq m-2$, we get $\beta_q(\mathcal{P}) = m-2$. 
  \item Otherwise, $H_i \neq H_3 \cap H_4$ or $H_i \subsetneq H_3 \cap H_4$ for any $i \in [1:2]$. For these cases, when we hit $H_i$, we invoke Theorem~\ref{theorem:look-ahead-skip} (cases~\ref{case1} or \ref{case2}) to avoid skipping further messages. So, $\beta_q \geq m-1$, which is achievable.
    \end{itemize}
\item If $L_\text{max}=3$, we must have the configuration $H_1 \subsetneq H_2 \subsetneq H_3$ and $H_3 \cup H_4 = [1:m]$, meaning that $(H_3,H_4)$ is not a nested pair.
  \begin{itemize}
  \item If $(H_1, H_3, H_4)$ or $(H_2, H_3, H_4)$ forms a perfect 2-nested absent receiver. Using the same argument above, we get $\beta_q \leq m-2$. Note that to get a lower bound of $\beta_q \geq m-3$, we must hit three absent receivers. This is not possible as upon hitting $H_1$, we can always skip a message $a \notin H_2$, and in doing so, we can only hit $H_3$ or $H_4$. Since $H_3 \cup H_4 = [1:m]$, we will not hit another absent receiver. So, $\beta_q \geq m-2$. This gives $\beta_q = m-2$.
  \item Otherwise, $H_1 \neq H_3 \cap H_4$ and $H_2 \neq H_3 \cap H_4$.
    \begin{itemize}
    \item If the first absent receiver to be hit is $H_2$, we can use Theorem~\ref{theorem:look-ahead-skip} (cases~\ref{case1} or \ref{case2}) to get $\beta_q \geq m-1$, which is achievable.
    \item Otherwise, we first hit $H_1$. We then use Theorem~\ref{theorem:look-ahead-skip} (cases~\ref{case1} or \ref{case2}). For case~\ref{case1} (that is, $H_1 \nsubseteq H_3 \cap H_4$), we immediately get $\beta_q = m-1$.  For case~\ref{case2} (that is, $H_1 \subsetneq H_3 \cap H_4$), following the proof of Theorem~\ref{theorem:look-ahead-skip}, we will skip a message $a$ in either $H_3$ or $H_4$. In any case, we can choose $a \notin H_2$. This choice allows us not to skip any more message. So, $\beta_q \geq m-1$, which is achievable.
    \end{itemize}
  \end{itemize}
  \end{itemize}
  \end{IEEEproof}
\fi

\end{document}